\documentclass{PoS}

\title{Internal shocks model for microquasar jets}

\ShortTitle{Internal shocks model for microquasar jets}

\author{\speaker{Omar Jamil}\thanks{Grateful to STFC for the Ph.D. funding}\\
        University Of Southampton\\
        E-mail: \email{o.jamil@phys.soton.ac.uk}}

\author{Rob Fender, Christian Kaiser\\
        University Of Southampton\\
	}

\abstract{We present an internal shocks model to investigate particle
  acceleration and radiation production in microquasar jets. The jet
  is modelled with discrete ejecta at various time intervals. These
  ejecta (or 'shells') may have different properties including the
  bulk velocity. Faster shells can catch up and collide with the
  slower ones, thus giving rise to shocks. The particles are
  accelerated inside the shocked plasma. Each collision results in a new
  shell, which may take part in any subsequent collisions as well as
  radiate due to synchrotron radiation. Almost
  continuous energy dissipation along the jet can be obtained with a
  large number of shell collisions. We
  investigate the spectral energy distribution of such jets as well as
  the physical significance of various parameters (e.g. the time
  interval between ejections and the shell size)}

\FullConference{VII Microquasar Workshop: Microquasars and Beyond\\
		 September 1-5 2008\\
		 Foca, Izmir, Turkey}

\begin{document}

\section{Introduction}
Jets are observed from a large variety of astronomical objects, ranging
from young stellar objects (YSO) to active galactic nuclei (AGN). In
terms of emission, flat radio ($\alpha = 0$ when $F_{\nu} \propto
\nu^{\alpha}$) spectra have commonly been observed from
these sources (AGN and X-ray binaries (XRB)). Observations of XRBs
suggest a frequency independent flat spectrum extending down to millimetre
\cite{fender00} or even near-infrared bands \cite{corbel02}. Flat radio
spectra from AGN and XRBs have traditionally been attributed to
partially self-absorbed jet emission from the innermost region of a
conical jet \cite{blandford79}, \cite{hjellming88}. However, in
order to obtain such flat spectra from partially self absorbed synchrotron
emission we either need continuous re-acceleration all along the emitting
jet region or use a non conical jet model \cite{kaiser06}. If we
assume a conical geometry for jets then the acceleration mechanisms
problem needs to be addressed. We explore internal shocks as a re-acceleration
mechanism for microquasar jets. 

\section{The model}
Our model has been inspired by the internal shocks model for
radio-loud quasars \cite{spada01}. We assume a conical geometry for
the jet with a given full opening angle $\theta_e$. Plasma is injected into
the jet in the form of discrete packets or ``shells''.

\subsection{Shell properties}
Our model is heavily based on Spada et al. model \cite{spada01} with a few
changes in order to adapt the model to XRBs. The injected shell
properties determine most of the observables for
the simulation. The central engine for shell ejections is considered
to be intermittent in nature, with a time interval ($\triangle_t$)
between ejections. Each shell also has a given bulk Lorentz
factor (BLF), $\Gamma$, associated with it. Outside of the collisions,
the $\Gamma$ does not evolve as
a shell moves through the jet. In order to choose $\Gamma$ for a given
shell we set the minimum($\Gamma_{min}\gtrsim 1$) and
maximum($\Gamma_{max}$) $\Gamma$ and then sample randomly from that range.

The shell mass is determined by the combination of kinetic luminosity of the jet
($L_W$), the time jet is on for ($t_{jet}$) and the total number of
shells to be injected during that time ($N$). The following relation
sets the total mass in the jet:

\begin{equation}
\sum_{j=1}^{N}M_j \Gamma_j c^2 = L_W t_{jet} \ ,
\label{eqn:totalMass}
\end{equation}
where $c$ is the velocity of light and $M_j$ is the mass of an arbitrary
shell. However, as we have assumed the central ejection engine to be 
intermittent in nature and accumulative, the time gap between ejections simply
accumulates more mass for a given shell.

\begin{equation}
M_j = \frac{L_W \triangle t_j}{\Gamma_j c^2} \ ,
\label{eqn:shMass}
\end{equation}  
where $\triangle t_j$ denotes the time interval between ejections $j-1$ and
$j$. We also set the width of the shell ($\triangle _j$).

In order to calculate the time gap between ejections, we take a
Gaussian distribution with the mean corresponding to the mean time gap
($\triangle t_j$) we desire. We then sample from that distribution
with chosen standard deviation. If the standard deviation is set to be
low then the average time gap between ejections will be close to the mean and
the total number of shells to be injected will be $\approx
t_{jet}/\triangle_t$. With a larger standard deviation, the total number
shells to injected will deviate from the previously stated relation. The variations in the
ejection time gap as well as the bulk Lorentz factors of the shells leads to
various shells catching up with each other, causing collision. 

\subsection{Two-shell collisions}
\label{sec:tshcol}
Once the above parameters are determined, we can set the injection times
as well as all the properties of the shells at the start of the
simulation. With the variations in $\Gamma$ and $\triangle t_j$ we can
have a large number of collisions taking place. The time gap between any two collision is:

\begin{equation}
dt_{col} =
\frac{R_{(j-1)}^{inner}-R_{j}^{outer}}{[\beta_{(j-1)}-\beta_{j}]c +
  0.5[\beta_{(j-1)}^{e}+\beta_{j}^{e}]c} \ ,
\label{eqn:colTime}
\end{equation}
where $R^{inner}$ and $R^{outer}$ are the radii to shell inner (with respect to
the source) and outer boundary i.e. the shell has a finite
width. $\beta$ is the shell velocity and
$\beta^e$ is the shell thermal expansion velocity, 

\begin{equation}
\beta^e = \frac{2\beta'_s}{\Gamma^2}\frac{1}{1-(\beta\beta'_s)^2} \ ,
\label{eqn:betaExp}
\end{equation}
where $\beta_s c = \upsilon'_s$ is velocity of sound in the plasma in the shell
co-moving frame with

\begin{equation}
\upsilon_s = \sqrt{\frac{1}{3}\frac{E'_{th}}{M}} \ ,
\label{eqn:soundV}
\end{equation}
$E_{th}$ denotes the thermal energy of the shell, causing the shell
to expand in between collisions.

The collisions are treated as inelastic collision. Momentum and
energy conservation laws give us the merged
shell properties. Merged shell mass is simply given by:

\begin{equation}
M_m = M_i + M_o \ ,
\label{eqn:mergMass}
\end{equation} 
where subscripts $i$ and $o$ denote inner faster and outer slower
shell involved in the collision. If we have the interacting shells bulk Lorentz
factor, Mass and internal energy ($\eta$), the merged shell bulk
Lorentz factor has the form:

\begin{equation}
\Gamma_m = \left(\frac{\mu_i\Gamma_i + \mu_o\Gamma_o}{\mu_i/\Gamma_i +
  \mu_o/\Gamma_o}\right)^2 \ ,
\label{eqn:blfMerg}
\end{equation}
where $\mu_i = M_i + \eta_i/c^2$ and $\mu_o = M_o + \eta_o/c^2$. The
internal energy of the merged shell is given by:

\begin{equation}
E_{in} = \eta_i + \eta_o + \mu_i c^2(\Gamma_i-\Gamma_m) + \mu_o
c^2(\Gamma_o - \Gamma_m).
\label{eqn:intMerg}
\end{equation}
The merged shell width is calculated by hydrodynamical treatment of the
shocked plasma. Adapting the jump equations \cite{blandford76} for
forward and reverse shock at the point of collision of the two shells
we get the width (in the lab frame) of the outer shock as:

\begin{equation}
\triangle_{sh,o} = \frac{\triangle_o}{\rho_o} \ ,
\label{eqn:shWidth}
\end{equation}
where $\triangle_o$ is the outer shell width and

\begin{equation}
\rho_o =
\frac{\Gamma_m}{\Gamma_o}\frac{\hat{\gamma}\Gamma_o+1}{\hat{\gamma}-1}
\ ,
\label{eqn:rhoOu}
\end{equation}
where $\hat{\gamma}$ is the adiabatic index. Equivalent equations
stand for the reverse shock involving the inner shell. The width of
the two corresponding shocks then gives the width of the newly formed
shell. It should be noted that the overall effect of the above
treatment is to cause compression in the plasma involved in the
collision.

\subsection{Shocked plasma}
The energy generated in the shocked compressed plasma is used in the
following three ways (assuming equipartition):

\begin{enumerate}
\item Accelerate electrons to a powerlaw distribution.
\item Generate magnetic field for the synchrotron radiation.
\item Thermal expansion of the shells. However, we do not explicitly take
  thermal electron distribution into account.    
\end{enumerate}
The energy split between the above is parameterized in the form of
$u_e$, $u_B$ and $u_{th}$. These determine the fraction of the total
energy generated that can be made available for the respective
process (i.e $u_e + u_B + u_{th} = E_{int}$). The parameter $u_{th}$ determines the
thermal energy content of a given shell which in turn determines the
local sound speed. This is used to calculate the
longitudinal (along the jet) expansion velocity of a shell. 

\subsection{Powerlaw electron distribution}
With each collision, the electrons are accelerated to a powerlaw
distribution of the form:

\begin{equation}
N(E)\textrm{d}E = \kappa E^{-p}\textrm{d}E \ ,
\label{eqn:plaw} 
\end{equation}
where $E = \gamma m_ec^2$ is the electron energy and $\kappa$ is the normalization
constant (which can be obtained by integrating the above relation
between $E_{min}$ and $E_{max}$). The acceleration is treated as instantaneous and uniform in
the respective volume. At each collision even the powerlaw
distribution of electrons is completely replaced by a new
one. However, in between collisions the powerlaw distribution ``ages''
via the evolution of $\kappa$ as well as $E_{max}$.

With the assumption of a conical jet and expanding shells (see
\ref{sec:tshcol}), we need to take adiabatic losses into account as
well (see \ref{sec:adloss}). 

\subsection{Magnetic field}
Parameter $u_B$ determines the fraction of energy available for the magnetic
energy ($E_B = u_B . E_{int}$):

\begin{equation}
E_B = \frac{B^2}{2\mu_0} \ ,
\label{eqn:magField}
\end{equation}
where $\mu_0$ is the magnetic permeability. This gives the magnetic
field strength available to synchrotron radiation. The magnetic field
is assumed to be randomly oriented and tangled in the plasma. This
allows us to treat the magnetic field as an ultra-relativistic gas
($\hat{\gamma}$ = 4/3), giving a straightforward adiabatic
loss treatment for the magnetic field.

\subsection{Adiabatic losses}
\label{sec:adloss}
As we are dealing with expanding shells, we need to take the adiabatic
losses into account. The shells expand both laterally due to conical
jet as well as longitudinally due to thermal expansion of the
shells. Using the treatment outlined in \cite{longair94},
we get a relation for the change in energy due to adiabatic
losses

\begin{equation}
\frac{E}{E_0} = \left(\frac{V}{V_0}\right)^{-1/\hat{\gamma}} \ ,
\label{eqn:aloss}
\end{equation} 
This relation can therefore be used for both electron energy losses as
well as the magnetic energy losses. 

\subsection{Synchrotron spectrum}
\label{sec:synch}
The Synchrotron radiation treatment outlined in \cite{longair94} is
employed in our model. As noted earlier, we assume that any given shell moves with
constant velocity along the jet, unless it collides. For a shell with
bulk Lorentz factor $\Gamma$ and velocity
$\beta$ and the jet axis making an angle $\theta$ with the observer,
we get the Doppler factor of

\begin{equation}
\delta_\mp = \left[\Gamma(1\mp\beta cos\theta)\right]^{-1} \ ,
\label{eqn:doppler}
\end{equation} 
where $\mp$ are the receding and approaching jets respectively. This then leads to observable monochromatic intensity

\begin{equation}
I_\nu = \delta_\mp^3 \frac{J_{\nu}}{4\pi\chi_{\nu}} \left(1 -
e^{-\chi_{\nu}r}\right) \ ,
\label{eqn:inten}
\end{equation}
where $J_{\nu}$ is the emissivity per unit volume, $\chi_{\nu}$ is
the absorption coefficient and $r$ is the radius of the shell. The
frequency is measured in the rest frame of the jet material. It is
related to the observed frequency by $\nu_{obs} = \delta_{\mp}\nu$. With
the estimate of distance to the source and equation \ref{eqn:inten},
we can calculate the flux density for any given shell.

\section{Results}
As mentioned earlier, the Radio to IR flat spectrum from a conical jet is a
commonly accepted picture (for XRBs and AGN), yet not fully explained. Therefore, with
our model in its early stages of development, obtaining the flat spectrum is quite an interesting
diagnostic. However, it is highly non-trivial to obtain a realistic
flat spectrum from a conical jet without making certain assumptions
about the re-acceleration mechanism. We present some preliminary
results that show the self consistently calculated synchrotron
spectra obtained from a conical jet with internal shocks taking place
along the entire length of the jet.

\begin{figure}[hf]
\centering
\includegraphics[scale=0.6]{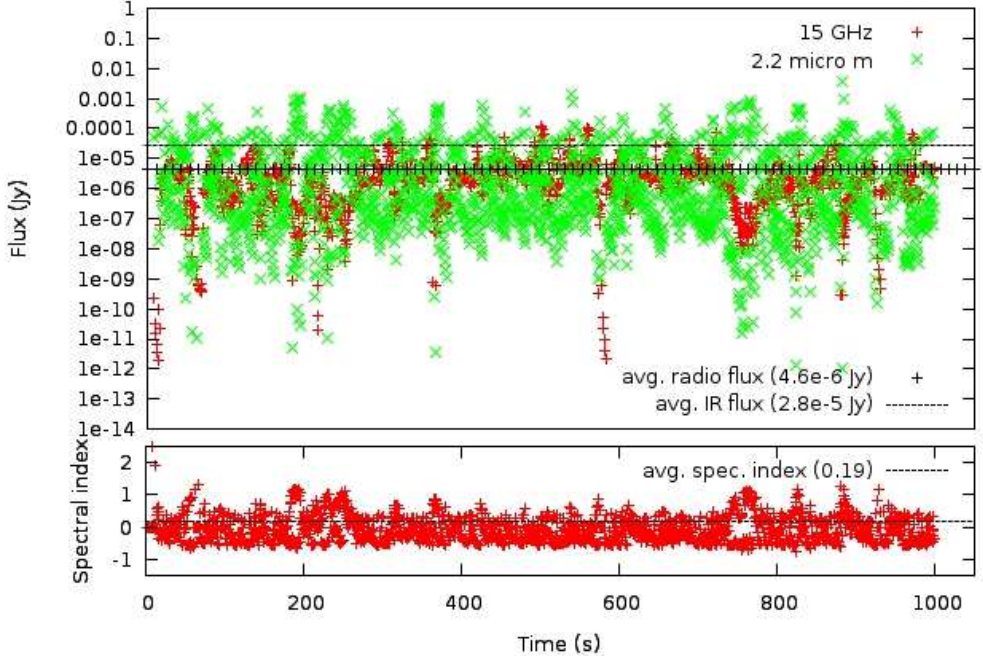}
\caption{Radio and infra-red lightcurve from the entire
  jet. The average fluxes and spectral index are shown with straight
  lines.}
\label{fig:figure1}
\end{figure}

\begin{figure}
\centering
\includegraphics[scale=0.6]{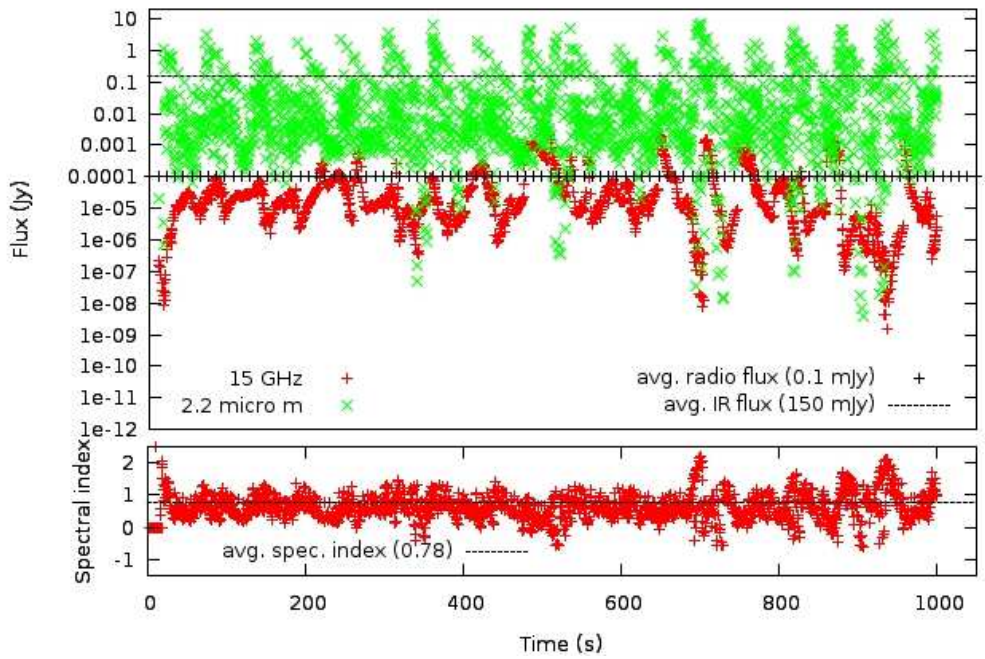}
\caption{Spectrum showing IR flux much greater than the Radio
  flux. Average flux and spectral index values are also shown.}
\label{fig:figure2}
\end{figure}
\begin{figure}
\centering
\includegraphics[scale=0.6]{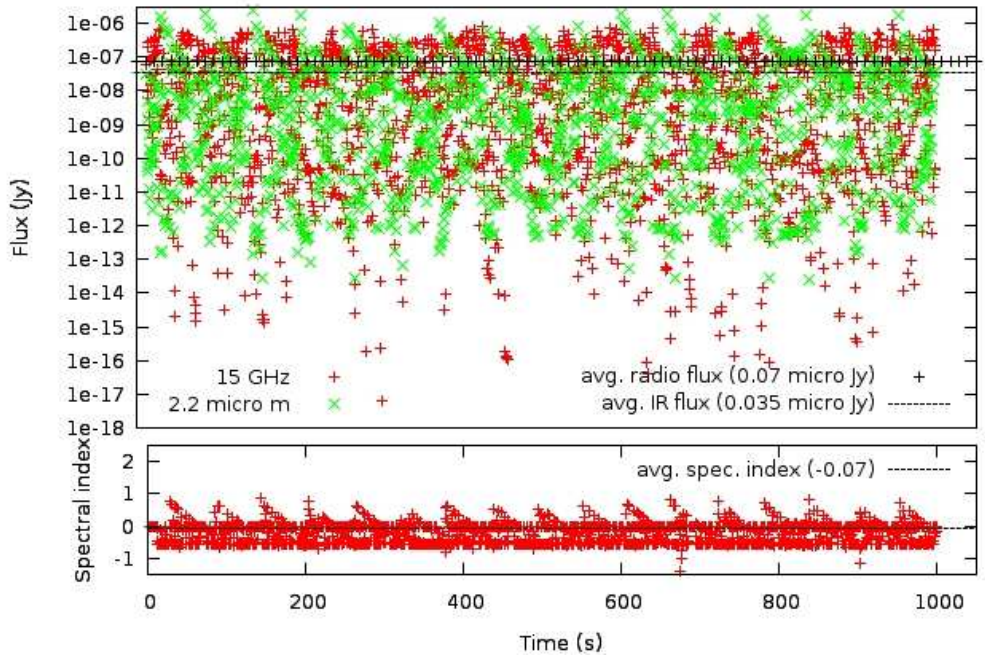}
\caption{A lightcurve form a weak jet. Average flux
  and spectral index values are also shown.}
\label{fig:figure3}
\end{figure}

Figure \ref{fig:figure1} shows the radio and
infra-red lightcurve from the entire jet. The lightcurves not only
shows how the infra-red and radio flux vary with time, but also the
spectral index. It can be seen that the spectral index oscillates
between $\alpha > 0$ (inverted) and $\alpha < 0$. The spectral index
can also be seen to spend time in the flat spectrum ($\alpha = 0$)
regime. Figure \ref{fig:figure2} shows the spectrum from jet a with
almost identical parameters as used for figure \ref{fig:figure1},
higher kinetic luminosity (the parameters are summarized in table \ref{table:results}). 
It can be seen in figure \ref{fig:figure2} that IR flux is much higher
than radio, giving rise to an overall inverted spectrum. The possible
reasons are discussed in the next section. For
completeness figure \ref{fig:figure3} shows a lightcurve from a
weak jet. This is primarily to show that with low
energy densities, the shell are almost instantly optically thin to
radio and hence the radio flux is greater than the IR flux. 

The key parameters giving rise to the spectra are summarized in table
 \ref{table:results}. The injection frequency is the average time gap between shell
injections and $\gamma_{max}$ is the integration upper limit for powerlaw
electron distribution (equation \ref{eqn:plaw}). Similarly $\Gamma_{max}$
is the upper limit for BLF sampling range.

\begin{table}
\caption{Model parameters to achieve the spectra shown (see Results
  section for details)}
\begin{center}
\begin{tabular}{llll}
\hline
Parameters & Figure 1 & Figure 2 & Figure 3 \\ \hline
Jet kinetic luminosity ($erg\ s^{-1}$) & $1\times10^{35}$ &$1\times10^{37}$ &$1\times10^{33}$\\ 
Distance ($kpc$) & 2 & 2 & 2\\
Injection freq. ($Hz$) & 1 & 1 & 1\\
$u_B$ & 0.1 & 0.1 & 0.1\\
$u_e$ & 0.1 & 0.1 & 0.1 \\
$u_{th}$ & 0.8 & 0.8 & 0.8\\
$\gamma_{max}$ (electrons)& $1\times 10^6$ & $1\times 10^6$ & $1\times 10^6$ \\
$\Gamma_{max}$ & 2 & 2 & 1.5 \\
Shell width ($m$) & $1\times 10^2$ & $1\times 10^2$ & $1\times 10^2$\\ \hline
\end{tabular}
\end{center}
\label{table:results}
\end{table}

\section{Discussion}
We saw in the results section that with exactly the same parameters
except the jet luminosity we get very different lightcurves. This can
can be explained by the fact that for the same number of shells,
increasing the jet kinetic luminosity increases the mass per
shell. This in turn has an influence on the energy generated at each
collision i.e. greater mass shell will have a larger energy
density. The energy density is the driving force behind the optical
depth for a given frequency in the shells and thus giving rise to a
different spectrum. In short, more massive shells remain optically thick to radio frequencies
for longer and thus completely changing the overall flux from the jet.

If we loook at figure \ref{fig:figure3}, we see that having a weaker jet
and thus lower energy density for each shell leads to most shells
becoming optically thin to radio very quickly and thus radio flux
overall being greater than the IR flux. This can be understood by looking
at the position of the self absorbed synchrotron spectrum peak relative to the
frequency in question \cite{longair94}.

For the most part $\Gamma_{max}$ was kept constant, except in the case
of spectrum in figure \ref{fig:figure3}. $\Gamma_{max}$, jet kinetic
luminosity and shell width have a similar effect on the overall
flux. The influence of these parameters is almost degenerate in the
sense that they influence the energy density of the shells. For
two colliding shells, greater the difference in their $\Gamma$s the
greater the amount of internal energy generated.

It should be noted that $u_{th}$ has been set high in the above
simulations in order to facilitate the longitudinal shell
expansion. This expansion should cause the shells to become optically
thin to radio quicker. We see that the fluxes shown in all the
spectra are not in very good agreement with the observations for
astrophysical systems with similar parameter values. In the case of
radio flux especially it is highly non-trivial to obtain a flat
spectrum with high fluxes. This indeed an interesting problem and as
noted earlier difficult to solve without making certain assumption
either about the re-acceleration mechanism \cite{blandford79} or the
jet geometry \cite{kaiser06}. In our model, with many collisions
taking place all along the jet, the shells are continually compressed
and powerlaw distributions completely replaced thus becoming optically
thick to radio. Shells then have to go through adiabatic expansion in order
to become optically thin to radio. However, the adiabatic energy
losses mean that the radio flux is significantly lower. We are
currently exploring the parameter space that may lead to this somewhat fine
balance between adiabatic losses and high radio flux for a flat spectrum. 

Although it was not shown in these proceedings,
the ejection frequency of the shells also plays an important role in the
lightcurve and spectra obtained from the jet. The ejection frequency
set at 1 Hz seems to give reasonable results and we not that it is
approximately the break frequency in X-ray power spectra of black
holes in bright hard state (e.g. Cygnus X-1).

The flaring seen in both radio and IR lightcurves is influenced by
various parameters. The flaring behaviour is determined by the balance
between how quickly the shells lose their energy via adiabatic losses
and the re-energization/compression due to collisions. Rapid adiabatic
losses (i.e. larger fraction of energy given to $u_{th}$) causes
greater flaring in the lightcurve. 
 
\section{Future directions}
As mentioned in the previous section, our initial aim is to explore
the parameter space that gives rise to the Radio-IR flat
spectrum. This is an attempt to break the degeneracy
present in some of the parameters. Once we have a reasonable picture
of what conditions are required to give rise a flat spectrum, we can
look at how X-ray variability maybe the driving force behind the shell
ejection variability. We also aim to add other physical
process to the model such as synchrotron self Compton radiation and
radiative losses for the electrons. 

\bibliography{bibliography}
\bibliographystyle{PoS}


\end{document}